\documentclass[superscriptaddress,preprintnumbers,
amsmath,amssymb,prd,nofootinbib,twocolumn,10pt]{revtex4}

\usepackage{graphicx}
\usepackage{epstopdf}
\usepackage{dcolumn}
\usepackage{bm}
\usepackage{hyperref}
\usepackage{color}

\begin{document}


\def\mpl{M_{\rm Pl}}


\title{Unified Model of Chaotic Inflation and Dynamical Supersymmetry Breaking}


\author{Keisuke Harigaya}
\affiliation{Department of Physics, University of California, Berkeley, California 94720, USA}
\affiliation{Theoretical Physics Group, Lawrence Berkeley National Laboratory,
Berkeley, California 94720, USA}
\author{Kai Schmitz}
\affiliation{Max-Planck-Institut f\"ur Kernphysik (MPIK), 69117 Heidelberg, Germany}


\begin{abstract}
The large hierarchy between the Planck scale and the weak scale can
be explained by the dynamical breaking of supersymmetry in strongly coupled
gauge theories.
Similarly, the hierarchy between the Planck scale and the
energy scale of inflation may also originate from strong dynamics,
which dynamically generate the inflaton potential.
We present a model of the hidden sector which unifies these two ideas, i.e.,
in which the scales of inflation and supersymmetry
breaking are provided by the dynamics of the same gauge group.
The resultant inflation model is chaotic inflation with
a fractional power-law potential in accord with the upper
bound on the tensor-to-scalar ratio.
The supersymmetry breaking scale can be much smaller
than the inflation scale, so that the solution to the
large hierarchy problem of the weak scale remains intact.
As an intrinsic feature of our model, we find that 
the sgoldstino, which might disturb the inflationary dynamics,
is automatically stabilized during inflation by dynamically generated 
corrections in the strongly coupled sector.
This renders our model a field-theoretical
realization of what is sometimes referred to as sgoldstino-less inflation.
\end{abstract}


\date{\today}
\maketitle


\section{Introduction}


Cosmic inflation not only solves the flatness and horizon problems of big bang cosmology~\cite{Guth:1980zm,Kazanas:1980tx,Linde:1981mu,Albrecht:1982wi},
but also explains the origin of the primordial density fluctuations that
seed the large-scale structure of the universe~\cite{Mukhanov:1981xt,Hawking:1982cz,Starobinsky:1982ee,Guth:1982ec,Bardeen:1983qw}.
To satisfy the upper bound on the tensor-to-scalar ratio in the power spectrum
of the cosmic microwave background (CMB)~\cite{Ade:2015lrj}, the potential
energy during inflation must be much smaller than the scale of gravity,
$\Lambda_{\rm inf} = V^{1/4} \lesssim 10^{-2} \mpl$.
The smallness of the energy scale of inflation, $\Lambda_{\rm inf}$,
is nicely explained
if the inflaton potential $V$ is generated by means of dimensional transmutation
in a strongly coupled gauge theory.
Refs.~\cite{Dimopoulos:1997fv,Izawa:1997df,Izawa:1997jc,Hamaguchi:2008uy} and \cite{Harigaya:2012pg,Harigaya:2014sua,Harigaya:2014wta,Harigaya:2014ola}
proposed models of small-field and large-field inflation along this idea, respectively.  


The electroweak scale also suffers from a hierarchy problem, $v_{\rm ew} \ll M_{\rm Pl}$,
which can be solved by supersymmetry and its breaking at a low energy scale~\cite{MaianiLecture,Veltman:1980mj,Witten:1981nf,Kaul:1981wp}.
Again, a plausible explanation for the smallness of the supersymmetry breaking
scale, $\Lambda_{\rm SUSY} \ll M_{\rm Pl}$, would be to presume that supersymmetry
is broken dynamically by strong dynamics~\cite{Witten:1981nf}.
So far, no evidence for superpartners of the standard model particles
has been found at the LHC, which has brought about the 
\textit{little} hierarchy problem, $v_{\rm ew} \ll m_{\rm SUSY}$ (where $m_{\rm SUSY}$
denotes a typical soft superparticle mass).
But supersymmetry nonetheless solves the \textit{large} hierarchy
problem, predicts the unification of the standard model gauge couplings and
provides a particle candidate for dark matter.
For these reasons, we take up the attitude that supersymmetry
as well as its dynamical breaking
are some of the leading candidates for new physics beyond the standard model.


In this letter, we propose a model of the hidden sector which unifies
these two ideas of dynamically generated energy scales.
The model resembles that of Refs.~\cite{Harigaya:2012pg,Harigaya:2014sua,Harigaya:2014wta}
during inflation; but the potential energy is non-zero even
after the end of inflation, which breaks supersymmetry.
The inflationary dynamics are those of chaotic inflation~\cite{Linde:1983gd}
with a fractional power-law potential.
The model is thus free from an initial conditions problem; and it is consistent
with the recent PLANCK data~\cite{Ade:2015lrj}.
See Refs.~\cite{Silverstein:2008sg,Takahashi:2010ky} for other
models of chaotic inflation with fractional power-law potentials.
We also refer to Ref.~\cite{Izawa:1997jc} for an earlier proposal
for the unified and dynamical generation of the energy scales of inflation
and supersymmetry breaking, which results in a scenario of
hybrid inflation~\cite{Linde:1991km,Linde:1993cn}.
This work has been followed up more recently in
Refs.~\cite{Schmitz:2016kyr,Domcke:2017xvu}, where it is demonstrated
how the dynamical breaking of supersymmetry at a very high energy scale
may result in scenarios of F-term and D-term inflation, respectively.
Finally, we refer to Ref.~\cite{Argurio:2017joe}, which considers
a perturbative model (as opposed to our strongly coupled models)
in which the inflaton potential as well as the breaking
of supersymmetry are both provided by the F term of a single chiral field.


\section{Dynamical chaotic inflation}
\label{sec:DCI}


We first review the idea of dynamical chaotic inflation (DCI) proposed in Refs.~\cite{Harigaya:2012pg,Harigaya:2014sua,Harigaya:2014wta}.
We start from a strongly coupled gauge theory which
generates a potential energy proportional to some power of the dynamical scale $\Lambda$,
\begin{align}
\label{eq:Vdyn}
V_{\rm dyn} \propto \Lambda^n.
\end{align}
To this theory, we add a pair of particles, $q$ and $\bar{q}$, that obtain their mass
from a coupling to the inflaton field $\phi$,
\begin{align}
{\cal L} = \lambda \phi\, q \bar{q}.
\end{align}
For a large field value of the inflaton, such that $\lambda \phi \gg \Lambda$,
the fields $q$ and $\bar{q}$ decouple; and around the dynamical scale the potential
energy in Eq.~(\ref{eq:Vdyn}) is generated.
Since the energy scale at which $q \bar{q}$ decouples depends on the inflaton field value, the dynamical scale also depends on it, through the running of the gauge coupling constant,
\begin{align}
\frac{\rm d}{{\rm dln}\, \mu} \frac{8\pi^2}{g^2(\mu)} = b,
\end{align}
where $\mu$ is the renormalization scale.
We shall denote the beta function coefficient $b$ in the high/low-energy theory
with/without $q \bar{q}$ as $b_{\rm HE}$ and $b_{\rm LE}$, respectively.
Then the effective dynamical scale $\Lambda (\lambda \phi)$ follows from
\begin{align}
\frac{8\pi^2}{g^2(\mu_0)} - \frac{8\pi^2}{g^2(\lambda \phi)} = & b_{\rm HE}\,
{\rm ln}\,\frac{\mu_0}{ \lambda \phi}\nonumber \\
\frac{8\pi^2}{g^2(\lambda \phi)} =& b_{\rm LE}\,
{\rm ln}\,\frac{\lambda \phi}{ \Lambda (\lambda \phi) },
\end{align}
where $g$ formally diverges, $g\left(\Lambda\right) \rightarrow \infty$,
at the dynamical scale.
Matching the running of the gauge coupling constant
at the $q \bar{q}$ mass threshold,
we obtain the dependence
\begin{align}
\label{eq:Lambda}
\Lambda \propto \phi^{\left(b_{\rm LE}-b_{\rm HE}\right)/ b_{\rm LE}}.
\end{align}
Together with Eq.~\eqref{eq:Vdyn}, this results in
a power-law potential for the inflaton, $\phi^p$, with the power
$p$ given as
\begin{align}
\label{eq:p}
p = n\,\frac{b_{\rm LE} - b_{\rm HE}}{b_{\rm LE}}.
\end{align}
This potential is suitable for inflation at large values
of the inflaton field, $\phi\gg \mpl$, which is nothing but a (dynamical)
realization of the idea of chaotic inflation.


The implementation of the above scheme into
supersymmetric theories is straightforward.
We start from a model of dynamical supersymmetry breaking,
add chiral multiplets $q$ and $\bar{q}$, and couple these chiral multiplets
to the inflaton multiplet $\Phi$.
To avoid the eta problem in supergravity~\cite{Ovrut:1983my,Holman:1984yj,Goncharov:1983mw,Coughlan:1984yk}
for a large field value of the inflaton, we introduce an
approximate shift symmetry $\Phi \rightarrow \Phi + i C$ in the K\"ahler potential~\cite{Kawasaki:2000yn,Kawasaki:2000ws}.
The negative contribution to the potential energy is suppressed
as long as the supersymmetry-breaking (Polonyi) field
has a field value much smaller than the Planck scale
during inflation. 


\section{Dynamical chaotic inflation and supersymmetry breaking unified}
\label{sec:DCI_SUSY}


In this section, we propose a model of dynamical chaotic inflation in which
the gauge dynamics also break supersymmetry in the true vacuum after the end of inflation.
The basic idea is the following:
We start from a dynamical supersymmetry breaking model
with a product group $G_1\times G_2$, such that supersymmetry is broken
by the strong dynamics of $G_2$, while the gauge interactions of $G_1$ merely lift
flat directions by a classical D-term potential.
To this model, we add $G_2$-charged matter fields $\ell$ and couple them
to an inflaton multiplet, $W = \lambda \Phi \ell^2$.
Supersymmetry is broken by the gauge dynamics of $G_2$ for large inflaton field values,
where the new matter multiplets decouple.
But for small field values, the gauge dynamics flow into a different phase;
and the potential energy proportional to the dynamical scale of $G_2$ and
hence the inflaton potential vanish.
By a suitable choice of matter fields and couplings, supersymmetry is instead now
broken by the strong dynamics of $G_1$ (or a subgroup of $G_1$, if the strong dynamics
of $G_2$ partially break $G_1$).
The supersymmetry breaking scale in the vacuum can be naturally much
smaller than the scale of inflation, provided there is a hierarchy between
the dynamical scales of $G_1$ and $G_2$ and/or the breaking of
supersymmetry by the strong dynamics of $G_1$ involves particularly
small couplings (realized, e.g., in the form of higher-dimensional operators).
In this paper, we shall present a simple realization
of this idea based on the groups $G_1= SU(5)$ and $G_2 = Sp(2)$.
Other examples will be given elsewhere.


\subsection{\texorpdfstring{\boldmath{$SU(5)\times Sp(2)$}}{SU(5) x Sp(2)} model during inflation}


Let us apply the idea described in Sec.~\ref{sec:DCI} to the
$SU(5)\times Sp(2)$ model of supersymmetry breaking~\cite{Dine:1995ag},
which is a generalization of the so-called $3$--$2$ model~\cite{Affleck:1984xz}.
The model is based on $SU(5) \times Sp(2)$ gauge dynamics and features chiral multiplets
$Q$, $\bar{U}$, $\bar{D}$ $L$, $\bar{q}_{1,2}$ in 
representations of the gauge group as listed in Tab.~\ref{tab:charge}.
Our convention for $Sp(N)$ groups is such that $Sp(1)\cong SU(2)$.
The theory contains the following flat directions,
\begin{align}
Q\bar{Q}L,~QQ\bar{Q}\bar{Q} \,,
\end{align}
where $\bar{Q} \in \left\{\bar{D},\bar{U},\bar{q}_i\right\}$.
The flat directions are lifted by introducing the following tree level superpotential,
\begin{align}
\label{eq:Wtree52}
W_{\rm tree} = y Q \bar{D} L + \frac{1}{M_*} Q Q \bar{q}_1\bar{q}_2 \,.
\end{align}


\begin{table}
\begin{center}
\begin{tabular}{|c||ccccc|}
\hline
 & $\quad Q \quad$ & $\quad \bar{D} \quad$ &  $\quad \bar{U} \quad$
 & $\quad \bar{q}_{1,2} \quad$ & $\quad L \quad$
\vphantom{$\Big[$} \\ \hline\hline
$\quad SU(5) \quad$ & ${\bf 5}$ & ${\bf \bar{5}}$ & ${\bf \bar{5}}$ &${\bf \bar{5}}$ & ${\bf 1}$
\vphantom{$\Big[$} \\
$\quad Sp(2) \quad$ &  ${\bf 4} $  & ${\bf 1}$  & ${\bf 1}$  & ${\bf 1}$ &  ${\bf 4} $ 
\vphantom{$\Big[$} \\ \hline
\end{tabular}
\caption{Matter content of the $SU(5)\times Sp(2)$ model.}
\label{tab:charge}
\end{center}
\end{table}


In this paper, we concentrate on the case where the dynamical
scale of $SU(5)$ is much smaller than that of $Sp(2)$, $\Lambda_{SU} \ll \Lambda_{Sp}$.
Supersymmetry is then broken by the deformed moduli constraint~\cite{Seiberg:1994bz}
of the $Sp(2)$ dynamics, which results in non-zero F terms for $\bar{D}$ and
the flat direction $QQ\bar{q}_1 \bar{q}_2$.
The potential energy is given by~\cite{Intriligator:1996pu}
\begin{align}
\label{eq:Vsp}
V_{Sp}\sim y^{3/2}  \left(\frac{\Lambda_{Sp}}{M_*}\right)^{1/2} \Lambda_{Sp}^4.
\end{align}
To turn this supersymmetry breaking model into a model
of dynamical chaotic inflation,
we add $Sp(2)$-charged chiral multiplets $\ell$ and
couple them to the inflaton field $\Phi$,
\begin{align}
W = \lambda \Phi \ell^2.
\end{align}
For $\lambda \Phi \gg \Lambda_{Sp}$ the extra multiplets $\ell$
decouple from the gauge dynamics.
The theory then exhibits supersymmetry breaking  and generates a non-zero potential energy.


The supersymmetry-breaking field is contained
in $\bar{D}$ and $QQ\bar{q}_1 \bar{q}_2$.
Its scalar component, the sgoldstino, is a flat direction
at tree level, which could potentially disturb the
inflationary dynamics.
It, however,
obtains a mass from strong-coupling corrections in the K\"ahler potential,
\begin{align}
m \sim y^{7/8}\frac{\Lambda_{Sp}^{9/8}}{M_*^{1/8}},
\end{align}
as is the case in generic models of dynamical supersymmetry breaking.
Unless $y$ is small, $m$ is much larger than the Hubble scale,
which provides a field-theoretical realization of
the so-called sgoldstino-less inflation~\cite{Ferrara:2014kva}.
This is a generic feature in models of dynamical chaotic inflation.
We note that the stabilization by a Hubble-induced mass would already
be enough to ignore the sgoldstino dynamics~\cite{Kawasaki:2000yn,Kawasaki:2000ws};
but the stabilization via IR quantum corrections
is advantageous in the sense that it is independent
of the unknown UV physics which determine the sign
and the magnitude of the Hubble-induced mass.


\subsection{Flow into \texorpdfstring{\boldmath{$SU(5)$}}{SU(5)} model in the vacuum}


After inflation, at  $\lambda \Phi \ll \Lambda_{Sp}$,
the extra multiplets $\ell$ no longer decouple, but participate
in the gauge interactions just like the other $Sp(2)$ flavors.
In Refs.~\cite{Harigaya:2012pg,Harigaya:2014sua,Harigaya:2014wta},
the fields $\ell$ as well as their couplings were
chosen so that the theory reaches a phase of s-confinement at low energies,
where all flat directions are lifted and supersymmetry is restored.
In this paper, we are instead going to chose the matter content and
couplings such that supersymmetry remains broken even in the true
vacuum after inflation.


We add a pair of $Sp(2)$ fundamentals, $\ell_1$ and $\ell_2$, and introduce
a coupling to the inflaton multiplet $\Phi$,
\begin{align}
\label{eq:Winf-ll}
W = \lambda  \Phi \ell_1 \ell_2 
\end{align}
The beta function coefficient of the $Sp(2)$ gauge coupling
at high and low energies is then given as $b_{\rm HE}=5$ and $b_{\rm LE}=6$,
respectively.
The potential energy during inflation scales like
$\Lambda_{Sp}$ to the power $n=9/2$, see Eq.~\eqref{eq:Vsp},
so that the exponent of the inflaton potential is given by $p=3/4$,
see Eq.~\eqref{eq:p}.
The dynamical scale around the vacuum, $\tilde{\Lambda}_{Sp}$,
and the dynamical scale during inflation $\Lambda_{Sp}$ are related to
each other as follows, see Eq.~\eqref{eq:Lambda},
\begin{align}
\Lambda_{Sp} = \tilde{\Lambda}_{Sp}
\left( \frac{\lambda \Phi}{ \tilde{\Lambda}_{Sp}} \right)^{1/6} \,.
\end{align}


Around $\Phi=0$, the $Sp(2)$ gauge theory reaches a phase of
s-confinement; and the low-energy theory is
described in terms of 28 gauge-invariant, composite meson fields,
\begin{align}
M_{QQ}\,,\: M_{QL}\,,\: M_{Q\ell_{1,2}} \,,\: M_{L\ell_{1,2}} \,,\: M_{\ell_1\ell_2} \,.
\end{align}
The fields ($M_{QL},\bar{D}$) and ($M_{\ell_1\ell_2},\Phi$)
obtain their masses from the superpotential in
Eqs.~(\ref{eq:Wtree52}) and (\ref{eq:Winf-ll}), respectively.
The inflaton mass around the origin is thus given by
\begin{align}
m_\Phi \sim \lambda \tilde{\Lambda}_{\rm Sp} \,.
\end{align}


After those fields decouple, the theory still contains the following chiral multiplets
\begin{align}
M_{QQ}\,({\bf 10}),\:
M_{Q\ell_{1,2}}\,({\bf 5}),\:
M_{L\ell_{1,2}}\,({\bf 1}),\:
\bar{U}\,({\bf \bar{5}}),\:
\bar{q}_{1,2}\,({\bf \bar{5}}),
\end{align}
where the numbers in bold refer to representations of $SU(5)$.
The superpotential in the s-confined phase reads
\begin{align}
W\sim& \frac{\tilde{\Lambda}_{\rm Sp}}{M_*} M_{QQ}\bar{q}_1\bar{q}_2 \nonumber \\
& +\frac{1}{\tilde{\Lambda}_{\rm Sp}}M_{QQ}^2
\left(  M_{Q\ell_1}M_{L\ell_2} + M_{Q\ell_2}M_{L\ell_1}  \right) \,.
\end{align}
Here, the second line is generated by the $Sp(2)$ dynamics.


The theory now contains one ${\bf 10}$, two ${\bf 5}$'s, and three ${\bf \bar{5}}$'s
of $SU(5)$.
By giving masses to two pairs of ${\bf 5} + {\bf \bar{5}}$, the theory
becomes nothing but the chiral supersymmetry breaking model
based on $SU(5)$, featuring one ${\bf 10}$ and one ${\bf \bar{5}}$
of $SU(5)$~\cite{Affleck:1983vc}.
The vacuum energy is then given by
\begin{align}
V_{\rm vac} \sim \tilde{\Lambda}_{SU}^4 \,,
\end{align}
where $\tilde{\Lambda}_{SU}$ is the dynamical scale of $SU(5)$
in the low-energy effective theory containing only ${\bf 10} + {\bf \bar{5}}$.
We may obtain a hierarchy between the inflation scale and the supersymmetry
breaking scale by choosing $\tilde{\Lambda}_{SU} \ll \Lambda_{Sp}$. 


The $SU(5)$ singlets $M_{L\ell_1}$ and $M_{L\ell_2}$ remain massless.
We can stabilize these fields by introducing $Sp(2)$ singlets
and coupling them to $L\ell_1$ and $L\ell_2$ in the quark picture at high energies.
Another possibility would be to simply introduce
a higher-dimensional operator, $W = L \ell_i L\ell_j$.


Depending on how we give masses to the two pairs of ${\bf 5} + {\bf \bar{5}}$,
the inflaton potential could be affected.
We may, e.g., remove the fields $M_{Q\ell_{1,2}}$ and $\bar{q}_{1,2}$
by adding the following superpotential in the quark picture,
\begin{align}
W = \kappa_1 Q \ell_1 \bar{q}_1 + \kappa_2 Q \ell_2 \bar{q}_2 \,,
\end{align}
such that the matter content of the $SU(5)$ supersymmetry breaking model
is provided by the chiral fields $M_{QQ}$ and $\bar{U}$.
After s-confinement of $Sp(2)$, those terms give masses to the
$(M_{Q\ell_1},\bar{q}_1)$ and $(M_{Q\ell_2},\bar{q}_2)$ pairs.
At the same time, during inflation and after
integrating out $\ell_1\ell_2$, this superpotential also
generates the second term in Eq.~(\ref{eq:Wtree52}) with $M_*\propto \Phi$.
When this inflaton-dependent term dominates over the $\Phi$-independent one,
the inflaton potential becomes the one with $p=3/4-1/2=1/4$.
If they are comparable to each other, we have
$p=1/4$ for small field values and $p=3/4$ for large field values.


\section{Phenomenology of inflation}


\subsection{CMB observables}


Taken all together, the model constructed in Sec.~\ref{sec:DCI_SUSY}
results in an inflaton potential of the following form,
\begin{align}
\label{eq:potential}
V = c\,\left|\frac{e^{i\alpha}}{M_*} + \frac{1}{\phi}\right|^{1/2}
\left(\frac{\lambda \phi}{\Lambda_{Sp}}\right)^{3/4} \Lambda_{Sp}^{9/2} \,.
\end{align}
Here, we choose a convention in which both $\phi$ and $M_*$
are real and positive; and the phase difference
between these two complex parameters is accounted for by the phase $\alpha$.
The parameter $c$ is a numerical constant, which we will set
to $c=1$ in the following.
The scalar potential is only monotonically increasing 
for positive $\phi$ as long as $\left|\alpha / \pi\right| \leq 5/6$.
For values of $\left|\alpha / \pi\right|$ closer to unity,
the potential exhibits a false vacuum at small field values.

\begin{figure}
\begin{center}
\includegraphics[width=0.95\columnwidth]{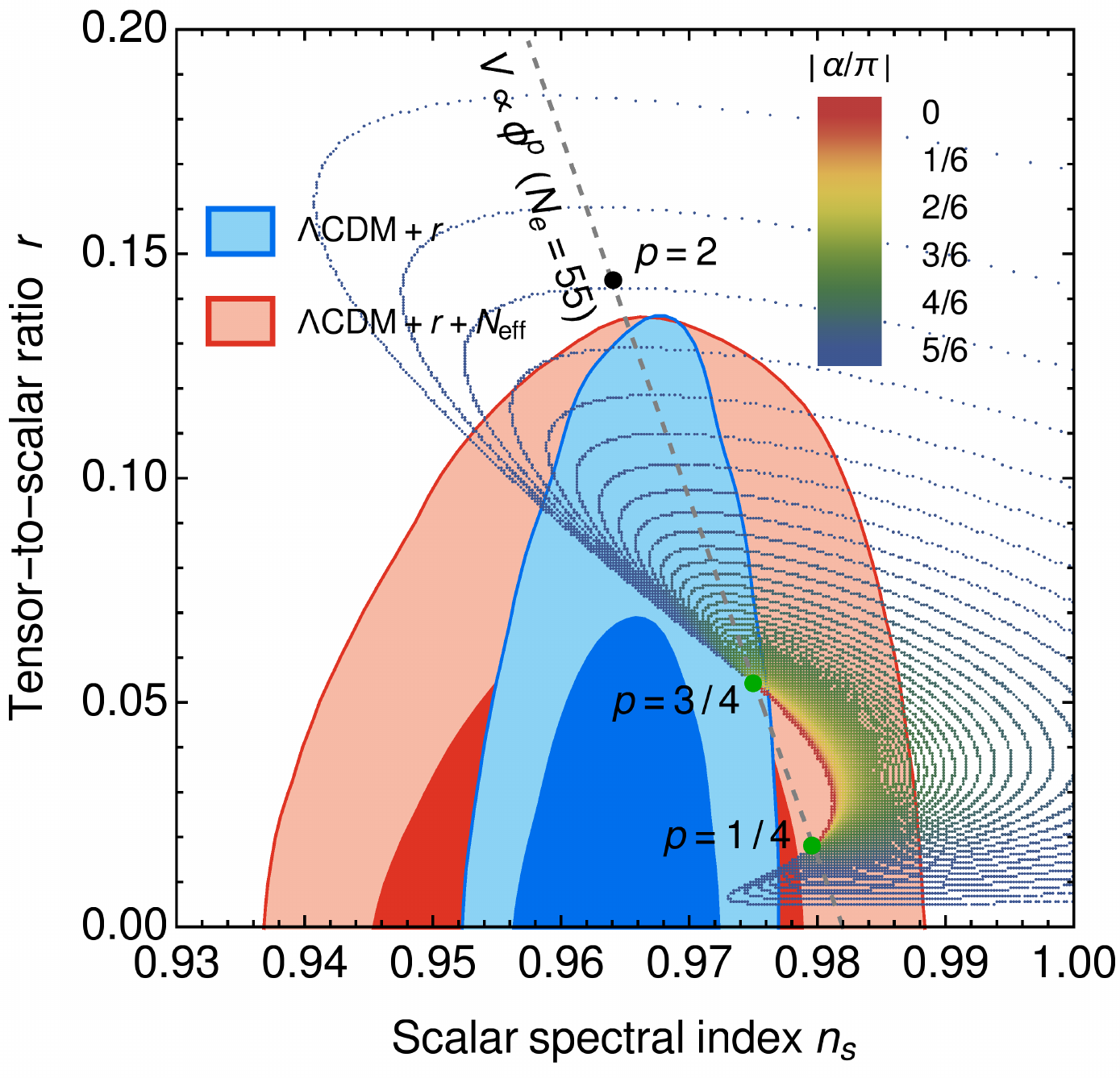}
\caption{Predictions of our model for $n_s$ and $r$,
compared with the latest constraints according to the PLANCK 2015 data
($68\,\%$ and $95\,\%$ C.\,L., TT,\,TE,\,EE\,$+$\,lowP)~\cite{Ade:2015lrj}.
The blue contours correspond to the standard $\Lambda$CDM\,$+$\,$r$ fit, 
whereas the red contours also take into account the possibility of a non-standard
number of relativistic degrees of freedom, $N_{\rm eff}$, at the time of photon decoupling.
The color scale indicates the value of the phase $\alpha$, which we vary on
a linear scale.
For each value of $\alpha$, we vary the mass scale $M_*$
in the interval $\left[10^{-3},10^3\right]M_{\rm Pl}$ on a logarithmic scale.
This results in \textit{orbits} in the $n_s$--$r$ plane that
smoothly connect the predictions of the pure power-law potentials
$\phi^{3/4}$ and $\phi^{1/4}$.
The local density of points in the above plot can be regarded
as a measure for how ``generic'' or ``typical'' a certain prediction is.
A low density of points indicates a rather special parameter choice,
while a high density of points indicates that a prediction is stable
under small variations of the input parameters $M_*$ and $\alpha$.}
\label{fig:cmb}
\end{center}
\end{figure}

From the potential in Eq.~\eqref{eq:potential}, we derive
the predictions for the CMB observables, i.e., for the scalar spectral index $n_s$
as well as for the tensor-to-scalar ratio $r$.
The result of our analysis is shown in Fig.~\ref{fig:cmb}.
The predictions for both parameters only depend on $M_*$ and $\alpha$.
If there is a clear hierarchy between $M_*$ and $\phi$ for all times
during inflation,
we simply recover the predictions for chaotic inflation based
on a standard power-law potential, $V \propto \phi^p$,
\begin{align}
n_s & = 1 - \frac{p+2}{2N_e} = 1 -  0.025
\left(\frac{p+2}{3/4+2}\right)\left(\frac{55}{N_e}\right) \,, \\
r & = \frac{4 p}{N_e} = 0.055 \left(\frac{p}{3/4}\right)\left(\frac{55}{N_e}\right) \,,
\end{align}
where $N_e$ is the number of $e$-folds at the CMB pivot scale.
For $M_*^{-1} \gtrsim M_{\rm Pl}^{-1}$, the $M_*^{-1}$ term in Eq.~\eqref{eq:potential}
clearly dominates over the $\phi^{-1}$ term.
In this case, we effectively obtain $p = 3/4$.
On the other hand, if the $M_*^{-1}$ term should be suppressed by 
a small coupling in Eq.~\eqref{eq:Wtree52} or by an (approximate) symmetry,
such that $M_*^{-1} \lesssim 0.01 M_{\rm Pl}^{-1}$, 
it can be neglected throughout inflation and we can effectively work with $p=1/4$.
For intermediate values of $M_*$, the predictions for $n_s$ and $r$
are more complicated, as they become sensitive to the phase $\alpha$.
This is evident from Fig.~\ref{fig:cmb}, where we show the variation of $n_s$
and $r$ for different values of $\alpha$.
In particular, we observe how, for fixed $\alpha$, the variation of
$M_*$ results in \textit{orbits} in the $n_s$--$r$ plane that connect the predictions
for $p = 3/4$ and $p=1/4$.


The parametric freedom of our model makes it easy to
achieve consistency with the recent PLANCK data~\cite{Ade:2015lrj}.
Our model predicts values of $r$ in the $r \sim 0.01 \cdots 0.1$ range
and is, therefore, in accord with the current upper bound, $r \lesssim 0.1$.
In particular, close-to-maximal values of the phase, $\alpha \simeq 5/6 \pi$,
allow to achieve rather large values of $r$, which are going to be tested in future
CMB experiments.
Our model moreover prefers values of $n_s$ in the $n_s \sim 0.97 \cdots 0.99$
range, which is slightly above the current best-fit value, $n_s \simeq 0.965$.
It is however interesting to note that the data still admits such relatively
large values of $n_s$, if it is fit by a $\Lambda$CDM\,$+$\,$r$\,$+$\,$N_{\rm eff}$
model, which also accounts for the possibility of dark radiation.


For given values of $M_*$ and $\alpha$, the observed amplitude of the scalar
power spectrum, $A_s \simeq 2 \times 10^{-9}$, fixes the parameter combination
$\lambda^{1/5} \Lambda_{Sp}$ in Eq.~\eqref{eq:potential}.
We find that, in the entire parameter space of interest, this product
is required to take a value of around
$\lambda^{1/5} \Lambda_{Sp} \sim 10^{16}\,\textrm{GeV}$.
At the same time, $\lambda$ should not be too small, since otherwise
the matter fields $\ell_1$ and $\ell_2$ do not decouple for the entire
duration of inflation.
We demand that $\lambda \phi \gtrsim \Lambda_{Sp}$ at all times during
inflation, which roughly translates into $\lambda \gtrsim 10^{-2}$,
see Ref.~\cite{Harigaya:2014wta} for details.
Given this lower bound on $\lambda$, we then find that the required value of $\Lambda_{Sp}$
is always remarkably close to the scale of grand unification.


\subsection{Reheating}


After inflation, the energy density stored in the inflaton field must be transferred
into standard model particles.
In our model, the inflaton resides in the supersymmetry breaking sector,
such that it may dominantly decay into particles in this sector.
Those particles eventually decay into gravitinos,
which easily leads to an overproduction of gravitinos.
We can forbid the decay mode into the supersymmetry breaking sector by
symmetry arguments.
For example, we can impose the $Z_2$ symmetry shown in Table~\ref{tab:Z2}, under
which the inflaton is odd.
The particles in the $SU(5)$ model, $M_{QQ}$ and $\bar{U}$, are $Z_2$-even and, hence,
the inflaton does not decay into these states.
The other $Z_2$-odd particles obtain masses proportional to $\tilde{\Lambda}_{Sp}$.
If $\lambda$ is sufficiently small, the inflaton ends
up being the lightest particle in the supersymmetry breaking sector,
so that it does not decay into any particles in this sector.


\begin{table}
\begin{center}
\begin{tabular}{|c||c|c|}
\hline
 & $\:\: \Phi \:\:$ $\:\: \bar{D} \:\:$ $\:\: \bar{q}_1 \:\:$
   $\:\: L \:\:$ $\:\: \ell_1 \:\:$
 & $\:\: Q \:\:$ $\:\: \bar{U} \:\:$ $\:\: \bar{q}_2 \:\:$ $\:\: \ell_2 \:\:$
\vphantom{$\Big[$} \\ \hline\hline
$\quad Z_2 \quad$ & $-$ & $+$ 
\vphantom{$\Big[$} \\ \hline
\end{tabular}
\caption{Charges under the $Z_2$ symmetry that forbids
the decay of the inflaton into the supersymmetry breaking sector.}
\label{tab:Z2}
\end{center}
\end{table}


The $Z_2$ symmetry also forbids the operator
$QQ \bar{q}_1 \bar{q}_2$ in Eq.~\eqref{eq:Wtree52}.
Therefore, if the $Z_2$ is an \textit{exact} symmetry, the $M_*^{-1}$ term
in Eq.~\eqref{eq:potential} is actually no longer present.
In our analysis, this corresponds to taking the limit $M_* \rightarrow \infty$,
such that the scalar potential reduces to an exact power-law with $p=1/4$.
On the other hand, if the $Z_2$ is only an \textit{approximate}
symmetry, it only suppresses the $M_*^{-1}$ term to some degree.
In this case, we have to work with the full scalar potential in 
Eq.~\eqref{eq:potential} and the predictions for the CMB observables
depend on the exact hierarchy between $M_*^{-1}$ and $\phi^{-1}$,
as discussed in the previous section.


The inflaton can decay, e.g., via a coupling
to the Higgs multiplets $H_{u,d}$ in the minimal supersymmetric standard model,
$W = \epsilon\, \Phi H_u H_d$. 
In this case, the $\mu$ term is generated via $Z_2$ symmetry breaking.
We may also identify the $Z_2$ with $R$ parity
and introduce $W = \epsilon_i\, \Phi L_i H_u$,
where the $L_i$ denote the standard model lepton doublets~\cite{Murayama:2014saa}.


\section{Discussion}


In this letter, we presented a strongly coupled model 
of the hidden sector based on $SU(5)\times Sp(2)$ gauge dynamics.
Our model combines the ideas of dynamical supersymmetry breaking
and dynamical chaotic inflation and, hence, explains the hierarchy
between the scales of supersymmetry breaking, inflation, and gravity,
$\Lambda_{\rm SUSY} \ll \Lambda_{\rm inf} \ll M_{\rm Pl}$.
During inflation, supersymmetry is broken because of the $Sp(2)$ deformed
moduli constraint.
This results in an inflaton potential that interpolates
between the power-law potentials $\phi^{3/4}$ and $\phi^{1/4}$, see Fig.~\ref{fig:cmb}.
The pseudoflat sgoldstino direction is automatically stabilized
during inflation by dynamically generated corrections in the K\"ahler potential.
After inflation, the $Sp(2)$ sector reaches a phase of s-confinement
and supersymmetry is broken by the $SU(5)$ gauge interactions.
In fact, at low energies, our model reduces to the chiral $SU(5)$ model
of dynamical supersymmetry breaking.


In the $SU(5)$ model, some approximate global symmetries are believed
to be spontaneously broken, which results in the presence of (pseudo-) Nambu-Goldstone bosons.
These bosons obtain non-zero field values in the early universe and
may affect the cosmological history.
Among them, the $R$ axion is potentially dangerous, since it has a mass squared
of $\mathcal{O}(m_{3/2}\tilde{\Lambda}_{SU})$ through the explicit breaking
of $R$ symmetry~\cite{Bagger:1994hh} and because it dominantly decays into gravitinos.
The gravitino eventually decays into the lightest supersymmetric particle (LSP),
which may lead to its overproduction.
Assuming that the initial amplitude of the $R$ axion is as large as
$\tilde{\Lambda}_{SU}$, the LSP abundance is estimated as
\begin{align}
\frac{\rho_{\rm LSP}}{s} \sim m_{\rm LSP}\,
\frac{T_{\rm RH}}{\mpl} \left(\frac{m_{3/2}}{\mpl} \right)^{1/4},
\end{align}
where $s$ is the entropy density and $T_{\rm RH}$ the reheating temperature.
Here, we imposed the condition that the universe must reach
a flat Minkowski vacuum after inflation, $m_{3/2}\mpl \sim \tilde{\Lambda}_{SU}^2$.
Requiring that $\rho_{\rm LSP}/s<4\times 10^{-10}$~GeV, we obtain an upper bound on $T_{\rm RH}$,
\begin{align}
T_{\rm RH} \lesssim 10^9\,\textrm{GeV} \left(\frac{100\,\textrm{TeV}}{m_{3/2}}\right)^{1/4}
\left(\frac{1\,\textrm{TeV}}{m_{\rm LSP}}\right) \,.
\end{align}


In the $SU(5)$ model, the gaugino masses of the minimal supersymmetric
standard model are generated only via anomaly
mediation~\cite{Randall:1998uk,Giudice:1998xp,Bagger:1999rd,
Boyda:2001nh,DEramo:2013dzi,Harigaya:2014sfa},
meaning that they are loop-suppressed compared to the gravitino mass.
The scalar masses, on the other hand, follow from the tree-level
K\"ahler potential and are as large as (or larger than) the gravitino mass.
For $m_{3/2} \sim \mathcal{O}\left(100\cdots1000\right)\,\textrm{TeV}$,
our model is thus compatible with the scenario of high-scale supersymmetry
breaking~\cite{Giudice:1998xp,Wells:2003tf,Wells:2004di}, which has gained
considerable interest after the discovery of the standard model Higgs boson
with a mass of $126$ GeV~\cite{Hall:2011jd,Ibe:2011aa}.


By choosing a different gauge group, we may also obtain a model of gauge mediation.
For example, we can modify our model by gauging only
the $SU(3)\times SU(2)\times U(1)$ subgroup of $SU(5)$.
By adding an appropriate superpotential term, supersymmetry is broken via
the $3$--$2$ model in the vacuum.
The $U(1)$ symmetry may be used as the messenger hypercharge~\cite{Dine:1994vc}.
We leave a detailed discussion of modifications of our model for future work.


\textit{Acknowledgments:}
The authors are grateful to the organizers of the 20th PLANCK conference
held at the University of Warsaw in May 2017, where this project was initiated. 
The work of K.\,H.\ was supported in part by the Director, Office of Science, and
Office of High Energy and Nuclear Physics of the U.\,S.\ Department of Energy under
contract DE-AC02-05CH11231 and by the National Science Foundation under grants
PHY-1316783 and PHY-1521446.
This project has received funding from the European Union's Horizon 2020
research and innovation programme under the Marie Sklodowska-Curie grant
agreement No.\ 674896 (K.\,S.).
K.\,S.\ acknowledges the hospitality of the
Laboratoire Univers et Particules de Montpellier 
during a stay at the University of Montpellier in July 2017,
where this project was finished.




\begin{thebibliography}{99} 


\bibitem{Guth:1980zm} 
  A.~H.~Guth,
  Phys.\ Rev.\ D {\bf 23}, 347 (1981).

\bibitem{Kazanas:1980tx} 
  D.~Kazanas,
  Astrophys.\ J.\  {\bf 241}, L59 (1980).

\bibitem{Linde:1981mu} 
  A.~D.~Linde,
  Phys.\ Lett.\  {\bf 108B}, 389 (1982).

\bibitem{Albrecht:1982wi} 
  A.~Albrecht and P.~J.~Steinhardt,
  Phys.\ Rev.\ Lett.\  {\bf 48}, 1220 (1982).

\bibitem{Mukhanov:1981xt} 
  V.~F.~Mukhanov and G.~V.~Chibisov,
  JETP Lett.\  {\bf 33}, 532 (1981)
  [Pisma Zh.\ Eksp.\ Teor.\ Fiz.\  {\bf 33}, 549 (1981)];
%
\bibitem{Hawking:1982cz} 
  S.~W.~Hawking,
  Phys.\ Lett.\ B {\bf 115}, 295 (1982).
  
\bibitem{Starobinsky:1982ee} 
  A.~A.~Starobinsky,
  Phys.\ Lett.\ B {\bf 117}, 175 (1982).
  
\bibitem{Guth:1982ec} 
  A.~H.~Guth and S.~Y.~Pi,
  Phys.\ Rev.\ Lett.\  {\bf 49}, 1110 (1982).
  
\bibitem{Bardeen:1983qw} 
  J.~M.~Bardeen, P.~J.~Steinhardt and M.~S.~Turner,
  Phys.\ Rev.\ D {\bf 28}, 679 (1983).

\bibitem{Ade:2015lrj} 
  P.~A.~R.~Ade {\it et al.} [Planck Collaboration],
  Astron.\ Astrophys.\  {\bf 594}, A20 (2016)
  [arXiv:1502.02114 [astro-ph.CO]].

\bibitem{Dimopoulos:1997fv} 
  S.~Dimopoulos, G.~R.~Dvali and R.~Rattazzi,
  Phys.\ Lett.\ B {\bf 410}, 119 (1997)
  [hep-ph/9705348].

\bibitem{Izawa:1997df} 
  K.~I.~Izawa, M.~Kawasaki and T.~Yanagida,
  Phys.\ Lett.\ B {\bf 411}, 249 (1997)
  [hep-ph/9707201].

\bibitem{Izawa:1997jc} 
  K.~I.~Izawa,
  Prog.\ Theor.\ Phys.\  {\bf 99}, 157 (1998)
  [hep-ph/9708315].

\bibitem{Hamaguchi:2008uy} 
  K.~Hamaguchi, K.-I.~Izawa and H.~Nakajima,
  Phys.\ Lett.\ B {\bf 662}, 208 (2008)
  [arXiv:0801.2204 [hep-ph]].

\bibitem{Harigaya:2012pg} 
  K.~Harigaya, M.~Ibe, K.~Schmitz and T.~T.~Yanagida,
  Phys.\ Lett.\ B {\bf 720}, 125 (2013)
  [arXiv:1211.6241 [hep-ph]].

\bibitem{Harigaya:2014sua} 
  K.~Harigaya, M.~Ibe, K.~Schmitz and T.~T.~Yanagida,
  Phys.\ Lett.\ B {\bf 733}, 283 (2014)
  [arXiv:1403.4536 [hep-ph]].

\bibitem{Harigaya:2014wta} 
  K.~Harigaya, M.~Ibe, K.~Schmitz and T.~T.~Yanagida,
  Phys.\ Rev.\ D {\bf 90}, 123524 (2014)
  [arXiv:1407.3084 [hep-ph]].

\bibitem{Harigaya:2014ola} 
  K.~Harigaya, M.~Ibe and T.~T.~Yanagida,
  Phys.\ Lett.\ B {\bf 739}, 352 (2014)
  [arXiv:1409.0330 [hep-ph]].

\bibitem{MaianiLecture}
L.~Maiani. in Proceedings: Summer School on Particle Physics, Paris,
	France (1979).
\bibitem{Veltman:1980mj} 
  M.~J.~G.~Veltman,
  Acta Phys.\ Polon.\ B {\bf 12}, 437 (1981).
  
\bibitem{Witten:1981nf} 
  E.~Witten,
  Nucl.\ Phys.\ B {\bf 188}, 513 (1981).
  
\bibitem{Kaul:1981wp} 
  R.~K.~Kaul,
  Phys.\ Lett.\ B {\bf 109}, 19 (1982).

\bibitem{Linde:1983gd} 
  A.~D.~Linde,
  Phys.\ Lett.\  {\bf 129B}, 177 (1983).

\bibitem{Silverstein:2008sg} 
  E.~Silverstein and A.~Westphal,
  Phys.\ Rev.\ D {\bf 78}, 106003 (2008)
  [arXiv:0803.3085 [hep-th]].

\bibitem{Takahashi:2010ky} 
  F.~Takahashi,
  Phys.\ Lett.\ B {\bf 693}, 140 (2010)\newline
  [arXiv:1006.2801 [hep-ph]].

\bibitem{Linde:1991km} 
  A.~D.~Linde,
  Phys.\ Lett.\ B {\bf 259}, 38 (1991).

\bibitem{Linde:1993cn} 
  A.~D.~Linde,
  Phys.\ Rev.\ D {\bf 49}, 748 (1994)
  [astro-ph/9307002].
  
\bibitem{Schmitz:2016kyr} 
  K.~Schmitz and T.~T.~Yanagida,
  Phys.\ Rev.\ D {\bf 94}, 074021 (2016)
  [arXiv:1604.04911 [hep-ph]].
  
\bibitem{Domcke:2017xvu} 
  V.~Domcke and K.~Schmitz,
  Phys.\ Rev.\ D {\bf 95}, 075020 (2017)
  [arXiv:1702.02173 [hep-ph]].

\bibitem{Argurio:2017joe} 
  R.~Argurio, D.~Coone, L.~Heurtier and A.~Mariotti,
  arXiv:1705.06788 [hep-th].
  
\bibitem{Ovrut:1983my} 
  B.~A.~Ovrut and P.~J.~Steinhardt,
  Phys.\ Lett.\  {\bf 133B}, 161 (1983).

\bibitem{Holman:1984yj} 
  R.~Holman, P.~Ramond and G.~G.~Ross,
  Phys.\ Lett.\  {\bf 137B}, 343 (1984).

\bibitem{Goncharov:1983mw} 
  A.~B.~Goncharov and A.~D.~Linde,
  Phys.\ Lett.\  {\bf 139B}, 27 (1984).

\bibitem{Coughlan:1984yk} 
  G.~D.~Coughlan, R.~Holman, P.~Ramond and G.~G.~Ross,
  Phys.\ Lett.\  {\bf 140B}, 44 (1984).

\bibitem{Kawasaki:2000yn} 
  M.~Kawasaki, M.~Yamaguchi and T.~Yanagida,
  Phys.\ Rev.\ Lett.\  {\bf 85}, 3572 (2000)
  [hep-ph/0004243].

\bibitem{Kawasaki:2000ws} 
  M.~Kawasaki, M.~Yamaguchi and T.~Yanagida,
  Phys.\ Rev.\ D {\bf 63}, 103514 (2001)
  [hep-ph/0011104].

\bibitem{Dine:1995ag} 
  M.~Dine, A.~E.~Nelson, Y.~Nir and Y.~Shirman,
  Phys.\ Rev.\ D {\bf 53}, 2658 (1996)
  [hep-ph/9507378].

\bibitem{Affleck:1984xz} 
  I.~Affleck, M.~Dine and N.~Seiberg,
  Nucl.\ Phys.\ B {\bf 256}, 557 (1985).

\bibitem{Seiberg:1994bz} 
  N.~Seiberg,
  Phys.\ Rev.\ D {\bf 49}, 6857 (1994)\newline
  [hep-th/9402044].

\bibitem{Intriligator:1996pu} 
  K.~A.~Intriligator and S.~D.~Thomas,
  Nucl.\ Phys.\ B {\bf 473}, 121 (1996)
  [hep-th/9603158].

\bibitem{Ferrara:2014kva} 
  S.~Ferrara, R.~Kallosh and A.~Linde,
  JHEP {\bf 1410}, 143 (2014)
  [arXiv:1408.4096 [hep-th]].

\bibitem{Affleck:1983vc} 
  I.~Affleck, M.~Dine and N.~Seiberg,
  Phys.\ Lett.\  {\bf 137B}, 187 (1984).

\bibitem{Murayama:2014saa} 
  H.~Murayama, K.~Nakayama, F.~Takahashi and\newline T.~T.~Yanagida,
  Phys.\ Lett.\ B {\bf 738}, 196 (2014)\newline
  [arXiv:1404.3857 [hep-ph]].

\bibitem{Bagger:1994hh} 
  J.~Bagger, E.~Poppitz and L.~Randall,
  Nucl.\ Phys.\ B {\bf 426}, 3 (1994)
  [hep-ph/9405345].

\bibitem{Randall:1998uk} 
  L.~Randall and R.~Sundrum,
  Nucl.\ Phys.\ B {\bf 557}, 79 (1999)
  [hep-th/9810155].

\bibitem{Giudice:1998xp} 
  G.~F.~Giudice, M.~A.~Luty, H.~Murayama and R.~Rattazzi,
  JHEP {\bf 9812}, 027 (1998)
  [hep-ph/9810442].

\bibitem{Bagger:1999rd} 
  J.~A.~Bagger, T.~Moroi and E.~Poppitz,
  JHEP {\bf 0004}, 009 (2000)
  [hep-th/9911029].

\bibitem{Boyda:2001nh} 
  E.~Boyda, H.~Murayama and A.~Pierce,
  Phys.\ Rev.\ D {\bf 65}, 085028 (2002)
  [hep-ph/0107255].

\bibitem{DEramo:2013dzi} 
  F.~D'Eramo, J.~Thaler and Z.~Thomas,
  JHEP {\bf 1309}, 125 (2013)
  [arXiv:1307.3251 [hep-ph]].

\bibitem{Harigaya:2014sfa} 
  K.~Harigaya and M.~Ibe,
  Phys.\ Rev.\ D {\bf 90}, 085028 (2014)
  [arXiv:1409.5029 [hep-th]].

\bibitem{Wells:2003tf} 
  J.~D.~Wells,
  hep-ph/0306127.

\bibitem{Wells:2004di} 
  J.~D.~Wells,
  Phys.\ Rev.\ D {\bf 71}, 015013 (2005)
  [hep-ph/0411041].

\bibitem{Hall:2011jd} 
  L.~J.~Hall and Y.~Nomura,
  JHEP {\bf 1201}, 082 (2012)
  [arXiv:1111.4519 [hep-ph]].

\bibitem{Ibe:2011aa} 
  M.~Ibe and T.~T.~Yanagida,
  Phys.\ Lett.\ B {\bf 709}, 374 (2012)
  [arXiv:1112.2462 [hep-ph]].

\bibitem{Dine:1994vc} 
  M.~Dine, A.~E.~Nelson and Y.~Shirman,
  Phys.\ Rev.\ D {\bf 51}, 1362 (1995)
  [hep-ph/9408384].
  

\end{thebibliography}
\end{document}